\documentclass[12pt,letterpaper]{article}

\usepackage{cite}
\usepackage{textcomp}
\usepackage{xcolor}

\usepackage{geometry}
\usepackage[compact]{titlesec}

\usepackage{caption}
\usepackage[shortlabels]{enumitem} 
\usepackage[font+=footnotesize]{subcaption}
\usepackage{amsmath,epsfig,amsfonts,amssymb,graphicx,bm,amsbsy,amssymb,url,tabularx,latexsym}
\usepackage{graphicx}
\usepackage{verbatim}
\usepackage{algorithm, algorithmic}

\newcommand{\sspace}[0]{{\cal S}}

\newcommand{\lagrangian}[0]{{\cal L}}
\newcommand{\bfx}[0]{{\boldsymbol  x}}
\newcommand{\bfc}[0]{{\boldsymbol  c}}
\newcommand{\bfb}[0]{{\boldsymbol  b}}
\newcommand{\bff}[0]{{\boldsymbol  f}}
\newcommand{\bfw}[0]{{\boldsymbol  w}}
\newcommand{\bfh}[0]{{\boldsymbol  h}}
\newcommand{\bfzero}[0]{{\boldsymbol  0}}
\newcommand{\bflambda}[0]{{\boldsymbol  \lambda}}
\newcommand{\bfnu}[0]{{\boldsymbol  \nu}}

\def\BibTeX{{\rm B\kern-.05em{\sc i\kern-.025em b}\kern-.08em
    T\kern-.1667em\lower.7ex\hbox{E}\kern-.125emX}}
\begin{document}

\title{An Optimization Framework for Certain Separable Problems using Neural Networks \\
}

\author{Rohit Negi and Soummya Kar  \\
	Carnegie Mellon University  \\
	}
\date{}

\maketitle

\begin{abstract} This paper studies a class of parametric constrained optimization problems that are motivated by applications in real time applications. Under a parameter-separable problem structure that naturally arises in these applications, the paper proposes a two phase strategy, based on offline learning and online processing, to address these optimization problems on resource limited  devices. Specifically, by exploiting the separable structure, an iterative Alternating Direction Method of Multipliers (ADMM) based solution procedure is developed that enables the use of certain learning based function representations (learned offline but readily computable online) to reduce the overall online on-device implementation complexity. By carefully crafting the ADMM procedure, it is shown that even as the parameters vary, the corresponding instances of the parametric optimization problem may be solved by lightweight online computations in the  device with the assistance of a neural network co-processor. 

\end{abstract}


\vspace{6truemm}

\section{Introduction}
\label{sec:introduction}

Consider the constrained convex optimization problem 
\begin{align}
\min_{\bfx} & \ f_{0}(\bfx;\bfc),  \quad \mbox {such that (s.t.)} \qquad \qquad \qquad\label{eqn:generalproblem} \\
f_{m}(\bfx;A_m) \ & \leq b_m, \quad m=1,2,\ldots,M,\label{eqn:generalconstraints}
\end{align}
with $f_0,f_{m}$ being convex functions that depend on multi-dimensional parameters $\bfc$ and $A_m$ respectively. We allow domain restrictions of the form $\bfx \in {\cal X}$ in this and subsequent problems discussed, but do not explicitly write that in the minimization problem.
Solving this problem using a standard convex optimization solver results in optimal $\bfx^*=\bfx^*(\bfc,A_1,\ldots,A_m)$. If this represents a problem being solved in a  device, such as a communication device,
such a device needs to be powerful enough to run a convex solver, and most likely also
needs carefully tuned numerical optimization libraries loaded, to obtain the required solution accuracy.
For example, unconstrained minimization using Gradient Descent may be used to minimize the Lagrangian function \cite{boyd2004convex} of \eqref{eqn:generalproblem}, which involves
calculating derivatives of potentially (numerically) complex functions, and evaluating these functions. Note that even for scalar $x$, evaluating a function such as $f_0(x) = \log(1+e^x)$
and its derivative requires numerical computation libraries that can accurately approximate
logarithm and exponential functions (such as by using iterative Newton root finding \cite{rheinboldt1998methods}). Further, a domain restriction $\bfx \in {\cal X}$ will potentially make this task even more challenging.

This raises an intriguing question: Is it
possible to dispense with heavy-duty numerical libraries, relying instead on only simple addition and multiplication operations, but assuming there is
a {\em neural network co-processor} available that can map parameters to
functions useful for numerical optimization? The rationale is that deep
neural networks have been found to provide accurate representations of 
fairly complex functions. And yet, their regular architecture allows 
networks of millions of parameters to be trained {\em offline}, so that
a device with such a co-processor can quickly compute the (neural network
represented) function, without relying on specialized numerical libraries. In essence, the solutions of optimization problems for various parameter values would be done offline (using massive computing power), and only the results
of the optimization would be stored as neural network represented functions for online computation. This would allow devices that are unable to run
full-fledged numerical libraries to perform complex optimization
(aided by the neural network co-processor). 

An extreme, but naive, application of this idea would be to consider a \emph{One shot } approximate solution of the parametric optimization problem~\eqref{eqn:generalproblem}-\eqref{eqn:generalconstraints} by learning the function $\bfx^*=\bfx^*(\bfc,A_1,\ldots,A_m)$ that maps parameter instances directly to the solution $\bfx^*$. However, this One shot  approach may quickly become practically infeasible as the number of parameters scale up. In fact, in the sample  problems that we consider in  Section~\ref{sec:modelexamples}, the number of parameters may scale up with the number of users, such as in a power allocation problem for cellular users, or with the number of constraints, which could be very many. While neural network type approximators are empirically known to be able to approximate fairly complex functions, the complexity (e.g., the number of neurons) required to carry out an accurate approximation may grow exponentially with the number of input parameters, even for well behaved function classes such as the class of continuous functions (see~\cite{haykin1998neural,telgarsky2016benefits,liang2016deep}). Additionally, learning or training these approximators would require sampling solutions of the parametric optimization problem at different parameter instances that also grows exponentially in the parameter dimension. This renders the One shot  approximation practically infeasible. This curse of dimensionality may be alleviated under certain special scenarios (e.g., more restricted classes of functions with certain smoothness properties \cite{haykin1998neural}). Unfortunately, even under the simplest type of polyhedral constraints, the solution $\bfx^*=\bfx^*(\bfc,A_1,\ldots,A_m)$ could turn out to be a fairly complex mapping of the parameters, e.g., non-smooth. Thus, theoretical approximation guarantees, say with Rectifier Linear Unit (RELU) based neural networks \cite{liang2016deep},  may not be directly applicable. 

Therefore, rather than aiming for a One shot neural network approximation, in this paper, we carefully use neural network based representations as helper functions at certain stages of the optimization pipeline with the two-fold objective of using function approximators when the underlying function to be approximated is amenable to \emph{lower complexity} representation (i.e., when the underlying parameter space is low-dimensional) while simultaneously ensuring that we minimize the dependence on special purpose numerical libraries, and only rely on  elementary additions and multiplications for our optimization algorithm. The usefulness of this idea is demonstrated by considering a variety of problems that are typically solved by real time devices (see Sections~\ref{sec:modelexamples} and~\ref{sec:simulation}).

\section{{Separable parameters Model}}
\label{sec:model}

This paper will focus on a specific  convex optimization problem, which we call a {\em Separable parameters problem}. To the best of our knowledge, such problems have not been explicitly discussed in previous research.
\begin{align}
\min_{\bfx_1,\ldots,\bfx_N}  \sum_{n=1}^N \sum_{l=1}^L f_{n,0,l}(\bfx_n; \bfc_{n,l})  \quad \mbox{s.t.}  \qquad \qquad \label{eqn:separableparametersproblem} \\
\sum_{n=1}^N \sum_{l=1}^L \bff_{n,m,l}(\bfx_n; A_{n,m,l})  \leq \bfb_m, \quad m=1,2,\ldots,M,
\end{align}
where each $\bff_{n,m,l}, \ m \geq 1$ function is a vector-valued function whose entries are convex in the argument $\bfx_{n}$. Further, assume that the $f_{n,0,l}$ functions are all strictly convex, perhaps by adding a term  $\frac{1}{2}\delta ||\bfx_n||^2$ to each function for a small $\delta>0$. The parameters $\bfc_{n,l}$ and  $\bfb_{m}$ are vectors, while parameters
$A_{n,m,l}$ are matrices. Not all terms in the problem need to appear, since we can always choose certain functions $\bff_{n,m,l}$ to be identically zero.  By {\em separable parameters}, we refer to the fact that the parameters of the problem appear {\em separately} in individual terms in the sums in the problem. i.e., no single term depends on a large set of parameters - a fact that is crucial for the optimization method we propose in Section \ref{sec:nnoptimization} to work with low complexity. 

Denote by $||\mathbf{p}||_0$ as the dimensionality of a parameter $\mathbf{p}$. For purposes of complexity estimation, we assume that the dimensionalities of these parameters are bounded as $||\bfc_{n,l}||_0 \leq ||\bfx_n||_0$ and
$||A_{n,m,l}||_0 \leq ||\bfx_n||_0\times ||\bff_{n,m,l}||_0$, with
$||\bfb_{m}||_0=||\bff_{n,m,l}||_0$.  Section \ref{sec:modelexamples} describes a few sample optimization problems  that have separable parameters.

\subsection{Discussion}
\label{sec:modeldiscussion}
The optimal solution $\bfx_n^*, \ \forall n$ to the problem can be found by using standard convex solvers. Clearly, this solution depends on all the
parameters $\bfc_{n,l}$,  $\bfb_{m}$ and $A_{n,m,l}$. For example, convex duality can be used to form the Lagrangian,
\begin{align}
\lagrangian \  &  = \sum_{n=1}^N   \sum_{l=1}^L f_{n,0,l}(\bfx_n; \bfc_{n,l})
+  \sum_{m=1}^M \bflambda_m^T ( \nonumber \\
& \phantom{111111} \sum_{n=1}^N \sum_{l=1}^L \bff_{n,m,l}(\bfx_n; A_{n,m,l}) - \bfb_m) \nonumber \\
& = \sum_{n=1}^N \left( \sum_{l=1}^L f_{n,0,l}(\bfx_n; \bfc_{n,l})
+ \right. \nonumber \\
& \phantom{111111}\left.  \sum_{m=1}^M   \bflambda_m^T (\sum_{l=1}^L \bff_{n,m,l}(\bfx_n; A_{n,m,l}) - \frac{1}{N} \bfb_m) \right), \label{eqn:partialfull}
\end{align}
where $\bflambda_m \geq \bfzero, \ m=1,\ldots,M$ are the dual variables.
The dual function is then obtained by minimizing $\lagrangian$ over all primal
variables $\bfx_n$, resulting in values $\bfx_n^\dagger$, which are optimal for the given values of the dual variables $\bflambda_m$. 
The optimal solution $\bfx_n^*$ of problem \eqref{eqn:separableparametersproblem} is then obtained by maximizing the dual function over
the dual variables $\bflambda_m$, and setting the resulting $\bfx_n^\dagger$ corresponding to the optimized dual variables as $\bfx_n^*$.

Convex problems which are separable in variables (or where the Lagrangian is separable)  are related to but not the same\footnote{In problems that are separable in {\em variables}, a large number of parameters may appear in individual terms, which is not allowed in our Separable parameters problem. On the other hand, our problem does not require separability of variables, as seen by the $L$ terms sharing the same variable.} as problem \eqref{eqn:separableparametersproblem}. The former problems are popular  due to reduced computational
burden on the optimizer. In particular, \eqref{eqn:partialfull} shows that the minimization over $\lagrangian$ reduces  to
$N$ separate minimizations of the partial sums $\lagrangian_n$ (the term in the parenthesis in \eqref{eqn:partialfull}) over individual $\bfx_n$ to obtain the $\bfx_n^\dagger$, which is computationally cheaper to implement than minimizing simultaneously over all primal variables. However, from the perspective of using a neural network co-processor, each such optimal $\bfx_n^\dagger$ is a function of parameters $\bfc_{n,l}$,  $\bflambda_m$ and $A_{n,m,l}$ for that particular value of $n$. 
So, a function that maps from the
parameter space to $\bfx_n^\dagger$ (given those parameter values) will be a high-dimensional function,
depending on up to $\sum_{l=1}^L ||\bfc_{n,l}||_0+\sum_{m=1}^M ||\bflambda_m||_0+\sum_{m=1}^M \sum_{l=1}^L ||A_{n,m,l}||_0$ parameters.
Due to the high complexity of the required neural network to represent such a function, the co-processor idea  described in Section \ref{sec:introduction} is not directly applicable to the problem, as formulated above, except in the trivial case where  the same function $\bff_{n,m,l}$ and the same parameters $\bfc_{n,l},A_{n,m,l}$ occur for all $m,l$ for a given $n$. Section \ref{sec:separableproblem} shows  a simple trick to allow the use
of {\em low-dimensional} neural networks to solve \eqref{eqn:separableparametersproblem}, which is the primary contribution of this paper.

 \subsection{Examples}
\label{sec:modelexamples}

We list a few optimization problems of interest in real time devices that fall within the
formulation \eqref{eqn:separableparametersproblem}. 

\begin{itemize}[leftmargin=2.5mm]
\item {\bf Power Control for interference mitigation}: Assume that there are $N$ nearby cellular users, each with an adaptable transmit power $P_n, n=1,2,\ldots,N$, transmitting to their respective  base stations \cite{biton}, where the channel power gain between transmitter $n$ and base station  $m$ is $s_{n,m}$,
and thermal noise  is $N_0$. The signals must be received with a Signal-to-Interference-Noise-Ratio (SINR) of $\gamma_n$ for
transmitter $n$. The problem is to minimize a cost based on an arbitrary  convex cost functions $f_{n,0}$ as follows.
\begin{align}
\min_{P_n, \ \forall n}  \ \sum_{n=1}^N  f_{n,0}(P_n; c_{n})  \quad \mbox{s.t.} \qquad  \qquad \qquad  \nonumber \\
N_0 + \sum_{n\ne m} P_{n}s_{n,m} \leq \frac{1}{\gamma_m}P_m s_{m,m} , \quad m=1,2,\ldots,N. \nonumber
\end{align}
To compare against \eqref{eqn:separableparametersproblem}, here $N=M$ and $L=1$, so $l\equiv 1$. The objective function is
$f_{n,0,l} = f_{n,0}(P_n; c_{n})$ and the constraint function for $m\geq 1$ is $f_{n,m,l} = P_n A_{n,m} $, where  $A_{n,m} =s_{n,m}$ if $n \ne m$,
and $A_{m,m} =  -\frac{1}{\gamma_m} s_{m,m} $.
For future reference, we note that the {\em x0-Predictor} and {\em x1-Predictor} neural networks described in Section \ref{sec:nnoptimization} only use 2 and 3 scalar features as input, respectively, here.

\item {\bf Robust Beam-forming}: Assume that a cellular user $0$ equipped with a $K$-antenna array \cite{schen} is receiving a signal from its base station at an angle $\theta_0$, at received power $P_0$. There are $L$ interferers also being received, with interferer $l$ arriving at angle $\theta_l$ and received power $P_l$. Assume that the interferer angles are measured with an error bound of $\pm \varepsilon$, since presumably they may be located in other cells. The robust beam-forming problem is to choose the $K$ dimensional array filter vector $\bfw$, so as to maximize the 
array gain towards the user $0$ signal, while keeping the worst case received interference plus noise within a bound of unity. 
\begin{align}
\min_{\bfw}  \  \left( - \mbox{Real}[{\bfw}^H {\bfh}(\theta_0)] \right)  + \frac{1}{2} \delta ||\bfw||^2 \ \ \mbox{s.t.}  \nonumber \\
\sum_{l=1}^{L} \left( \max_{|e| \leq \varepsilon} |{\bfw}^H {\bfh}(\theta_l+e)|^2 \right) \frac{P_l}{N_0} +  ||\bfw||^2 \leq \frac{1}{N_0}. \quad \nonumber
\end{align}
Here, ${\bfh}(\theta)$ is the known antenna array response to a signal arriving at angle $\theta$ (which depends on the geometry of the array), 
$N_0$ is the noise power, and $(\cdot)^H$ denotes Hermitian.
To compare against \eqref{eqn:separableparametersproblem}, here $N=M=1$, $L$ is  number of interferers\footnote{While the above problem is
stated in terms of complex variables, we can easily reduce it to a real-valued problem by defining variable $\bfx$ as $(\mbox{Real}[\bfw^T],\mbox{Imag}[\bfw^T])^T$. It is a convex problem because the maximum of convex quadratic functions is convex.}. 
$f_{n,0,l} = - \frac{1}{L}\mbox{Real}[{\bfw_l}^H {\bfh}(\theta_0)]  +  \frac{1}{2L} \delta ||\bfw||^2$ and  $f_{n,1,l} =\left( \max_{|e| \leq \varepsilon} |{\bfw_l}^H {\bfh}(\theta_l+e)|^2 \right) \frac{P_l}{N_0} +   \frac{1}{L} ||\bfw_l||^2$.
Looking ahead in this case, the  {\em x1-Predictor} neural network described in Section \ref{sec:nnoptimization} uses the features $\theta_l, \lambda \frac{P_l}{N_0} , \frac{\lambda}{L}, \Delta \bfnu_{l}$, which can be problematic due to the potentially high-dimensional $\Delta \bfnu_{l}$. So, a second use of the formulation is required to minimize $\lagrangian_{n,1,l}$ in \eqref{eqn:xnmlopt},  using the fact that $\lagrangian_{n,1,l}$ here is the maximum of quadratic functions.

 \item {\bf Generalized waterfilling under uncertainty}: In an OFDMA system, each user is  allotted separate resources (called `resource blocks' in 5G cellular)  in downlink transmission, where each resource is a set of carriers and/or an antenna beam \cite{molisch} over one subframe. The transmit power in each resource can be adapted by the base station based on the channel power gain $s$ in that resource, to minimize a convex function of these powers under a  desired user bit rate  constraint. However, $s$ is expected to be erroneous due to channel state information (CSI) feedback based on noisy pilot reception. (For example, in 5G, CSI-Reference Signal \cite{parkvall} is transmitted in a few  sub-carriers, from which CSI of all sub-carriers is estimated using least squares estimation, using the fact that the channel impulse response is time-limited.) This can be approximately modeled as below. Suppose there are $N$ resources allotted to a user, with {\em nominal} power gains $s_n$ for the $n^{th}$ resource (as reported by the CSI based on measured reference signals ${\boldsymbol \theta}$), and the base station will allot a transmit power $P_n$ to that resource. Due to erroneous measurement, the actual gain over that subframe are random variables $(\tilde{s}_1,\ldots,\tilde{s}_N) \in \sspace({\boldsymbol \theta})$. Here,  $\sspace({\boldsymbol \theta})$ is the set of gains close to the nominal gains, but allowing for uncertainty in measured ${\boldsymbol \theta}$. We discretize $\sspace({\boldsymbol \theta})$ and write it as the finite set of sequences $\{(s_{1,m},\dots,s_{N,m}), m=1,\ldots,M\}$.   
 Then, the cost minimization problem under a {\em worst case} desired bit rate constraint $C$ is,
 \begin{align}
\min_{0 \leq P_1,\ldots,P_N \leq P_{max}}  \  \sum_{n=1}^N  f_{n,0}(P_n; c_{n})  \quad \mbox{s.t.}  \qquad  \label{eqn:waterfillingrobust}\\
\min_{(\tilde{s}_1,\ldots,\tilde{s}_N) \in {\cal S}({\boldsymbol \theta})} 
\left( \sum_{n=1}^N r(P_n; \tilde{s}_n) \right)  \geq C. \qquad \label{eqn:waterfillingrobustconstraint}
\end{align}
Here, the convex cost functions $f_{n,0}$ are arbitrary, such as $P_n+c_n P_n^2$ if high power is costly for certain sub-carriers (perhaps due to inter-cell interference considerations).  
The concave  bit rate function $r(P_n; \tilde{s}_n)$ is typically $\log_2(1+\tilde{s}_n \frac{P_n }{N_0} \gamma_c) $, with $\gamma_c$ being the loss due to an imperfect channel code. This problem is a generalization of the standard `waterfilling' problem in communications, since it allows arbitrary costs and  models gain uncertainty. The problem can be converted to the form \eqref{eqn:separableparametersproblem} by writing the constraints as $- \sum_{n=1}^N r(P_n; s_{n,m}) \leq -C, \ m=1,\ldots,M$. So, comparing this problem against \eqref{eqn:separableparametersproblem}, here $M=|\sspace|$ and $L=1$. 
$f_{n,0,l} = f_{n,0}(P_n; c_{n})$ and  for $m\geq 1$, $f_{n,m,l} = -r(P_n; s_{n,m})$.
The {\em x0-Predictor} and {\em x1-Predictor} neural networks described in Section \ref{sec:nnoptimization} only use 2 and 3 scalar features as input, respectively, here.
\end{itemize}

\begin{figure*}[t]
\centering
\begin{minipage}{.64\columnwidth}
\centering
\includegraphics[width=0.99\textwidth]{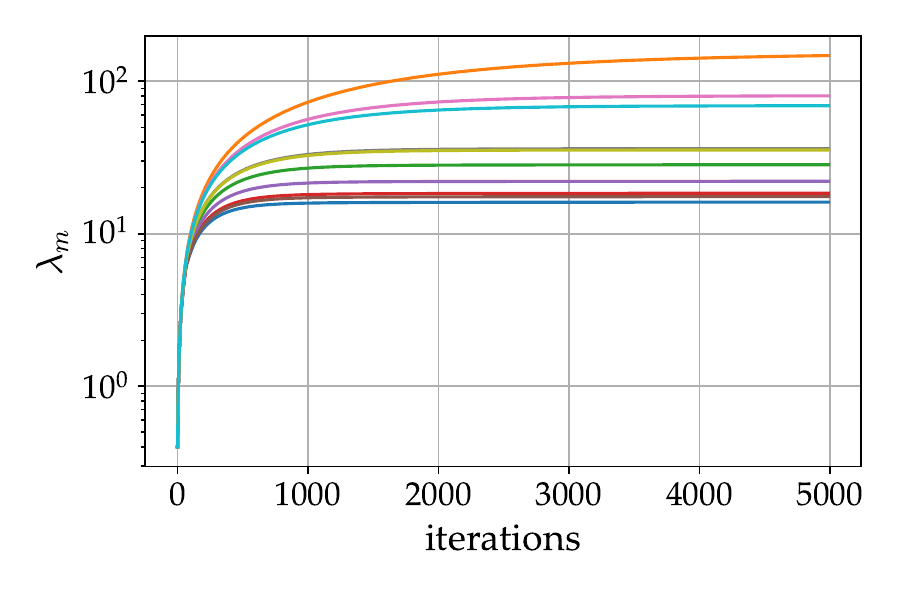}
\end{minipage}
\hfill
\begin{minipage}{.64\columnwidth}
\centering
\includegraphics[width=0.99\textwidth]{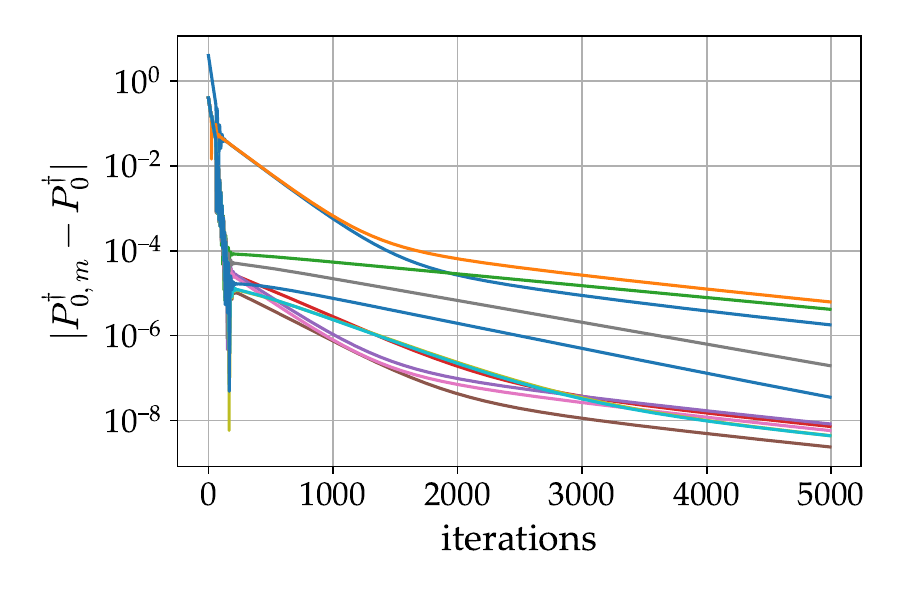}
\end{minipage}
\hfill
\begin{minipage}{.64\columnwidth}
\centering
\includegraphics[width=0.99\textwidth]{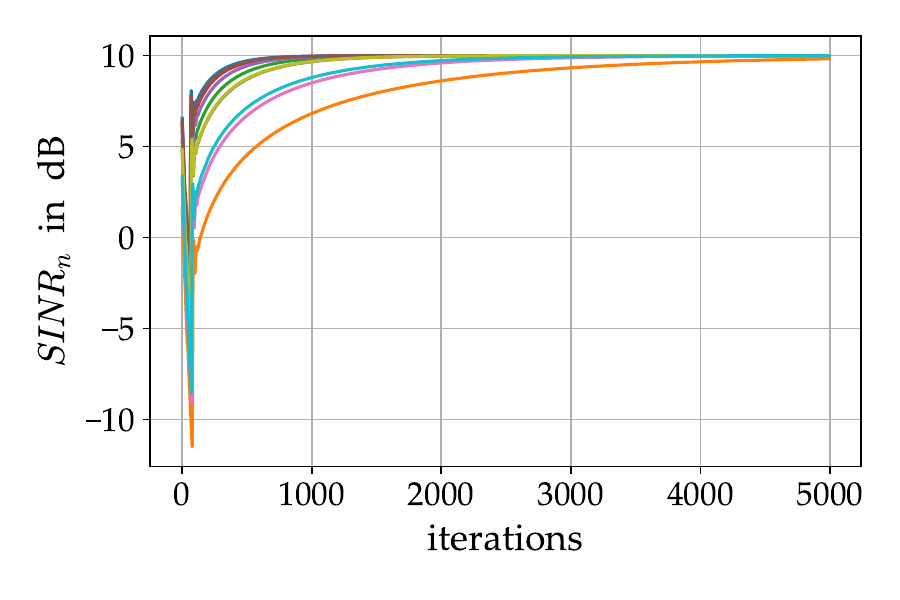}
\end{minipage}
\caption{Convergence of (left to right): (a) $\lambda_m$,  (b) $(M+1)$ copies $P_{0,m}^\dag$ around $P_0^\dag$, and (c) $SINR$s.}
\label{fig:convergence}
\end{figure*}

\section{Duality using Multiple copies}
\label{sec:separableproblem}

For the Separable parameters problem \eqref{eqn:separableparametersproblem}, let's reformulate the problem as
the following equivalent problem. Make $L\times (M+1)$ additional copies $\bfx_{n,m,l}, m=0,1,\ldots,M, l=1,2,\ldots,L$ of each variable $\bfx_n$. In the subsequent discussion, the phrase $\forall m$ is short for $m=0,1,\ldots,M$. Then, the problem \eqref{eqn:separableparametersproblem} is equivalent to,
\begin{align}
\min_{\substack{\bfx_n, \forall n\\ \bfx_{n,m,l}, \ \forall n,m,l}} \ \  \sum_{n=1}^N \sum_{l=1}^L f_{n,0,l}(\bfx_{n,0,l}; \bfc_{n,l}) \quad +  \qquad \nonumber \\
\frac{\rho}{2}  \sum_{n=1}^N \sum_{m=0}^M  \sum_{l=1}^L ||\bfx_{n,m,l} - \bfx_n||^2  \qquad \mbox{s.t.}  \qquad \label{eqn:equivproblem}  \\
\sum_{n=1}^N \sum_{l=1}^L \bff_{n,m,l}(\bfx_{n,m,l}; A_{n,m,l}) \leq \bfb_m, \ \ m=1,\ldots,M, \\
\bfx_{n,m,l}  = \bfx_n, \qquad \forall n,m,l. \qquad  \qquad   \qquad   \label{eqn:primalequalities}
\end{align}
Here, the equalities \eqref{eqn:primalequalities} ensure that the copies $\bfx_{n,m,l}$ 
 are all equal to $\bfx_{n}$ for each $n$. Superfluous quadratic terms weighted by fixed $\rho>0$ are added in the objective, because
we will use the well known Alternating Direction Method of Multipliers (ADMM) \cite{boyd2011distributed} to optimize the obtained dual problem, since it is generally more stable than the simpler Dual Ascent method.
For this purpose, we calculate the Lagrangian of this reformulated problem,
\begin{align}
\lagrangian & =
   \sum_{n=1}^N \sum_{m=0}^M \sum_{l=1}^L \lagrangian_{n,m,l}(\bfx_{n,m,l}) - \sum_{m=1}^M \bflambda_m^T  \bfb_m \quad \nonumber \\
&    - \frac{\rho}{2} \sum_{n=1}^N \sum_{m=0}^M \sum_{l=1}^L ||\bfnu_{n,m,l}||^2, \quad  \mbox{where} \label{eqn:separatedseparableproblem}
\end{align}
\begin{align}
\lagrangian_{n,0,l}(\bfx_{n,0,l}) \! \doteq  \!
f_{n,0,l}(\bfx_{n,0,l}; \bfc_{n,l}) \! + \!  \frac{\rho}{2}  ||\bfx_{n,0,l}\! - \! \bfx_n\! +\! \bfnu_{n,0,l}||^2,  \ \  \nonumber
\end{align}
\begin{align}
\lagrangian_{n,m,l}(\bfx_{n,m,l})   \doteq  \bflambda_m^T \bff_{n,m,l}(\bfx_{n,m,l}; A_{n,m,l})   + \qquad \quad \nonumber \\
\frac{\rho}{2}  ||\bfx_{n,m,l} - \bfx_n +\bfnu_{n,m,l}||^2 ,  \ \ m=1,2,\ldots,M.  \  \label{eqn:partiallagrangian}
\end{align}
Here, each $\lagrangian_{n,m,l}$ is a partial-Lagrangian function over a single variable $\bfx_{n,m,l}$ but also over the variable $\bfx_n$. The $\rho \bfnu_{n,m,l}$, which are of the same dimensionality as $\bfx_{n,m,l}$, are the
(scaled) dual variables corresponding to \eqref{eqn:primalequalities}. 

Note that if the superfluous quadratic terms in the objective \eqref{eqn:equivproblem} are dispensed with, the Lagrangian $\lagrangian$ in \eqref{eqn:separatedseparableproblem} becomes separable (since the coupling variables $\bfx_n$ disappear); a standard primal-dual iterative update in this case reduces to the classical Dual Ascent, which is however slow to converge. We thus retain the quadratic terms, to essentially form the augmented Lagrangian, leading to an ADMM update rule (described below), which is known to be more stable and achieve faster convergence. The idea of ADMM is to handle the coupling variables $\bfx_n$ by minimizing $\lagrangian$ in \eqref{eqn:separatedseparableproblem} (which includes the quadratic) in  two steps. In the first step, $\lagrangian$  is
minimized only over the $\bfx_{n,m,l}$ variables, assuming that $\bfx_n$ variables and the dual variables are fixed. For this step, 
 \eqref{eqn:separatedseparableproblem} shows the $\lagrangian$ is separable in the variables $\bfx_{n,m,l}$, so that $N\times(M+1)\times L$ parallel minimizations can be done  to obtain,
\begin{align}
\bfx_{n,m,l}^\dag & \doteq \underset{\bfx_{n,m,l}}{\mbox{argmin}} \ 
\lagrangian_{n,m,l}(\bfx_{n,m,l}), ~~\forall n,m,l. \label{eqn:xnmlopt}
\end{align}
$\bfx_{n,m,l}^\dag $ is unique because
the quadratic terms in \eqref{eqn:partiallagrangian} are strictly convex, since $\rho>0$ (which also explains the stability of ADMM).
Define $\Delta \bfnu_{n,m,l} = \bfnu_{n,m,l}-\bfx_n$. Note that $\bfx_{n,0,l}^\dag $ depends only on the values of $\bfc_{n,l},\Delta \bfnu_{n,0,l}$, while for $m\geq 1$, $\bfx_{n,m,l}^\dag $ depends only on the values of $A_{n,m,l},\bflambda_m,\Delta \bfnu_{n,m,l}$.
In the second step, the minimization over each $\bfx_n$ can be done easily because it only appears in the quadratic terms, resulting in the unique closed-form optimal value,
\begin{align}
\bfx_n^\dag & = \frac{1}{(M+1)L} \sum_{m=0}^M \sum_{l=1}^L (\bfx_{n,m,l}^\dag +\bfnu_{n,m,l}), \ \forall n. \quad \label{eqn:xnopt}
\end{align}
With $\bfx_{n,m,l}^\dag $ and $\bfx_n^\dag$ obtained, the ADMM increases the dual function using the following dual variable updates for
an appropriate step size
$\mu>0$.
\begin{align}
\bflambda_m & \leftarrow \max\left(0, \bflambda_m + \mu(\sum_{n=1}^N \sum_{l=1}^L  \right. \nonumber \\
& \phantom{1111111} \left. \bff_{n,m,l}(\bfx_{n,m,l}^\dag; A_{n,m,l}) - \bfb_m) \right) \label{eqn:lambdaupdate} \\
\bfnu_{n,m,l} & \leftarrow  \bfnu_{n,m,l} +  \left(\bfx_{n,m,l}^\dag - \bfx_n^\dag \right). \label{eqn:nuupdate}
\end{align}
The values $\bfx_n^\dag$ calculated in \eqref{eqn:xnopt} are used as the known values of $\bfx_n$ for the first step in the next iteration. 
The algorithm is initialized with an appropriate initial choice of  $\bfx_n$ and dual variables $\bflambda_m,\bfnu_{n,m,l}$.
After the ADMM dual variable updates converge to the optimal values $\bflambda_m^*,\bfnu_{n,m,l}^*$,  standard
strong duality arguments show that the corresponding optimal $\bfx_n^\dag$
is the optimal solution $\bfx_{n}^*$ of the problem \eqref{eqn:equivproblem}, and thus, of the Separable parameters problem \eqref{eqn:separableparametersproblem}.

\begin{figure*}[t]
\centering
\begin{minipage}{.64\columnwidth}
\centering
\includegraphics[width=0.99\textwidth]{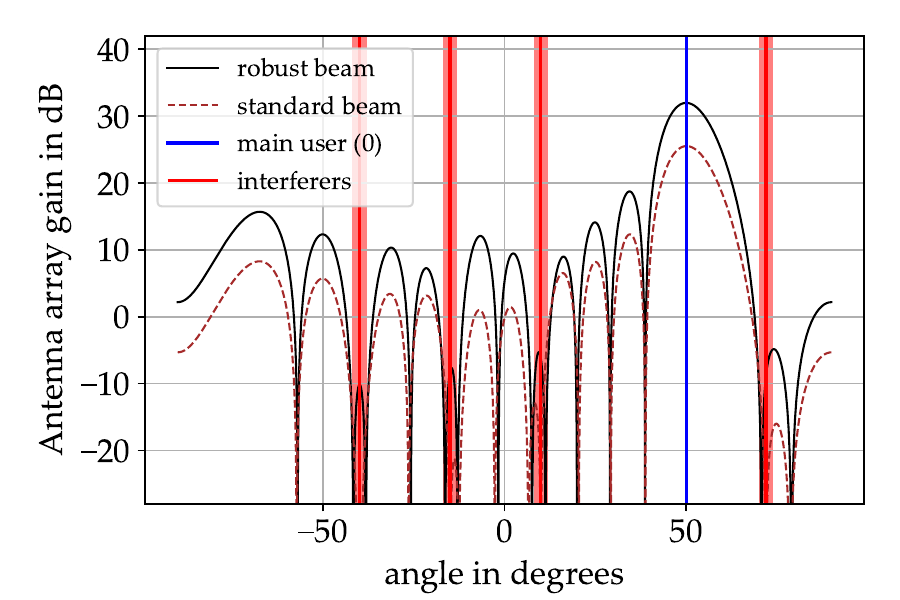}
\caption{Antenna array pattern.}
\label{fig:antennagain}
\end{minipage}
\hfill
\begin{minipage}{.64\columnwidth}
\centering
\includegraphics[width=0.99\textwidth]{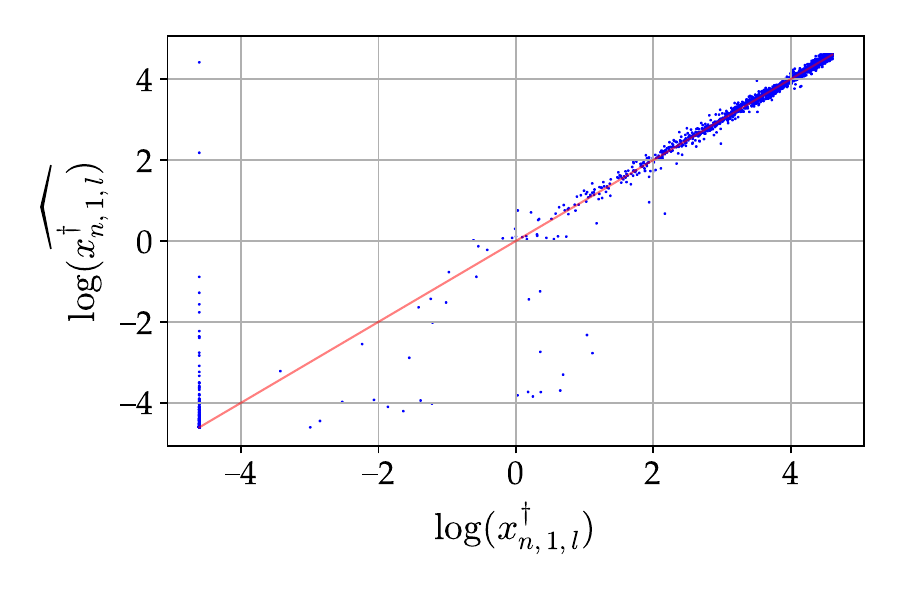}
\caption{{\em x1-Predictor} accuracy.}
\label{fig:xnml}
\end{minipage}
\hfill
\begin{minipage}{.64\columnwidth}
\centering
\includegraphics[width=0.99\textwidth]{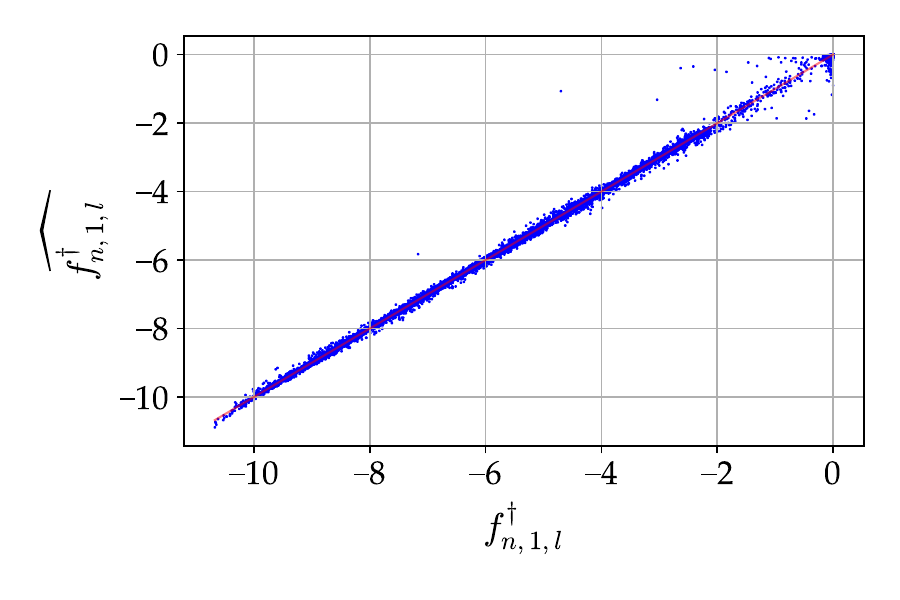}
\caption{{\em f1-Predictor} accuracy.}
\label{fig:fnml}
\end{minipage}
\end{figure*}

\section{Neural Network aided optimization}
\label{sec:nnoptimization}

The ADMM algorithm described in Section \ref{sec:separableproblem} suggests the following two phase method to solve the 
main problem \eqref{eqn:separableparametersproblem}, by solving the equivalent problem \eqref{eqn:equivproblem} using {\em Neural network representations}. 
\begin{enumerate}[label={\bf Phase \Roman*},wide=\parindent,labelindent=0em]
\item {\em (Offline)}: This offline phase has the following steps, which  can be done using large scale computing such as cloud computing, since it is not limited to using the computing power of the real time device.
\begin{enumerate}[i),leftmargin=1.5em]
\item Sample the parameters $\bfc_{n,l}, A_{n,m,l}, \bflambda_m, \Delta \bfnu_{n,m,l}$ over an appropriate range that covers practical use cases of the problem. For each sample $(\bfc,A,\bflambda, \Delta \bfnu)$,
solve the partial Lagrangian problems \eqref{eqn:xnmlopt}, to obtain the
optimal $\bfx_{n,m,l}^\dag$ and then evaluate the corresponding function $\bff_{n,m,l}^\dag \doteq  \bff_{n,m,l}(\bfx_{n,m,l}^\dag)$, for each $n,m,l$.   In the terminology of machine learning, this produces a {\em dataset} of potentially millions of data points; each with {\em features} $(\bfc,A,\bflambda,\Delta \bfnu)$ and {\em targets}  $\bfx_{n,m,l}^\dag$ or $\bff_{n,m,l}^\dag$.
\item From Section \ref{sec:separableproblem}, it is known that $\bfx_{n,0,l}^\dag,\bff_{n,0,l}^\dag$ are  functions of $\bfc, \Delta \bfnu$ and for $m\geq 1$, $\bfx_{n,m,l}^\dag,\bff_{n,m,l}^\dag$  are  functions of $A, \bflambda, \Delta \bfnu$.
So, using the above dataset, we train deep neural networks with sufficient number of layers and neurons per layer, using inputs (features) $(\bfc,\Delta \bfnu)$ 
to predict $\bfx_{n,0,l}^\dag,\bff_{n,0,l}^\dag$ and separate deep networks\footnote{Note that the neural networks are meant to {\em represent} the given functions $\bfx^\dag,\bff^\dag$,  by being trained to {\em predict} those functions.} using inputs  $(A,\bflambda,\Delta \bfnu)$ to predict $\bfx_{n,m,l}^\dag,\bff_{n,m,l}^\dag, m\geq 1$. We assume that these $\bfx^\dag$ and $\bff^\dag$ are predicted by separate networks, although a single network predicting both is also possible.

In several use cases, such as the problems described in Section \ref{sec:modelexamples}, the objective functions are identical, i.e., $\bff_{n,0,l}=\bff_0, \forall n,l$. Similarly, the constraint functions  are identical, $\bff_{n,m,l}=\bff_1, \forall m\geq 1, n,l$. For ease of exposition, we will restrict discussion below to such cases only, although the method is applicable to the general case also. In such  cases, we only need to obtain neural networks for two cases; $m=0$ and $m=1$, which we will call {\em x0-Predictor}, {\em x1-Predictor},   {\em f0-Predictor}, {\em f1-Predictor}
to represent $\bfx_{n,0,l}^\dag,\bfx_{n,1,l}^\dag,\bff_{n,0,l}^\dag,\bff_{n,1,l}^\dag$, respectively.
\end{enumerate}
\item ({\em Online}): This phase is run by the real time device solving the optimization problem \eqref{eqn:equivproblem} in realtime. Given the observed parameters
$\bfc_{n,l}, A_{n,m,l}, \bfb_m$ in the realtime problem, the ADMM algorithm \eqref{eqn:lambdaupdate}, \eqref{eqn:nuupdate} is run, beginning with appropriate initial
values of the dual variables $\bflambda_m,\bfnu_{n,m,l}$. \eqref{eqn:lambdaupdate} requires calculation of $\bff_{n,m,l}(\bfx_{n,m,l}^\dag; A_{n,m,l})$ for each $m\geq 1,n,l$, which can be directly obtained as the prediction
of neural network {\em f1-Predictor}. We also need to calculate $\bfx_{n,m,l}^\dag, m\geq 0,n,l$, both, for the update \eqref{eqn:nuupdate} as well as to calculate $\bfx_n^\dag $ using the averaging \eqref{eqn:xnopt}. This can be directly obtained as the prediction
of neural networks {\em x0-Predictor} and {\em x1-Predictor} for the parameters corresponding to each $n,m,l$. After ADMM converges, the final value of $\bfx_n^\dag $ is the solution to
the main problem
\eqref{eqn:separableparametersproblem}. The {\em f0-Predictor} is only needed to calculate the value of the optimized objective.
\end{enumerate}
Notice that the complexity of the critical {\bf Online Phase II} lies in calculation of the
$\bfx^\dag,\bff^\dag$ optimal values, which has been offloaded to a dedicated neural network co-processor in our framework. This leaves only the simple addition and multiplication operations of the ADMM \eqref{eqn:lambdaupdate}, \eqref{eqn:nuupdate} and of the averaging step \eqref{eqn:xnopt} (with $\frac{1}{(M+1)L}$ in the last part being pre-computed and stored in a table for various values of $M,L$). This dispenses of the need for a
specialized numerical computation library carefully designed for solving optimization problems, and offloads the complex issues of ensuring convergence of iterative minimization algorithms, such as enforcing domain restrictions $\bfx \in {\cal X}$ or choosing the correct step sizes, to the offline solution, which can be extensively tested to ensure that optimal solutions to \eqref{eqn:xnmlopt} are indeed being obtained in forming the dataset. Further, given the regular architecture of deep neural networks, we conjecture that it may be cheaper to use such a co-processor, instead of high-speed Digital Signal Processors capable of handling arbitrary computations, although we acknowledge that the benefit of this substitution needs to be carefully verified by a comprehensive comparison on various axes, such as cost, energy, speed, etc. Note that the neural networks used in realtime for inference (i.e., calculating $\bfx^\dag,\bff^\dag$) are only trained offline. Therefore, the co-processer, which only runs the neural network in evaluation mode, does not need to be powerful, unlike graphical processor units (GPUs) that are typically used to train such neural networks.

To compare the above approach against the more straight-forward approach of using a
neural network to directly represent the optimal solution $\bfx^* = \bfx^*(\bfc_{n,l}, A_{n,m,l}, \bfb_m \forall n,m,l)$ of \eqref{eqn:separableparametersproblem}, notice that the latter neural network, which we refer to as {\em One shot network}, must predict based on an input (feature) of dimensionality $\sum_{l=1}^L ||\bfc_{n,l}||_0+\sum_{m=1}^M ||\bfb_{m}||_0+\sum_{m=1}^M \sum_{l=1}^L ||A_{n,m,l}||_0$, which will likely need a prohibitively complex neural network due to the high input dimensionality. On the other hand, the neural networks in {\bf Online Phase II} of our formulation has input dimensionality upper bounded by
$||\bfc_{n,l}||_0+ ||A_{n,m,l}||_0+ ||\bflambda_{m}||_0+ ||\bfnu_{n,m,l}||_0$, which will typically be much lower. As an example, in \eqref{eqn:separableparametersproblem}, if there are $N=10$ variables $\bfx_n$, each of which is a scalar, $M=10$ inequalities of scalar valued functions $\bff$, and $L=10$, the One shot neural network must predict based on an input of 120 dimensions, while the {\em x0-Predictor} and {\em x1-Predictor} neural networks have an input of only 2 and 3 dimensions, respectively (which is the same for the  two {\em f-Predictors}.) Finally, it is also clear that the learned neural network in the One shot case requires making assumptions about the critical constants  $M$ and $L$ (for example, number of transmitters in the Power Control scenario, or number of interferers in the Beam-forming scenario in Section \ref{sec:modelexamples}), with a different network being learned for different values of $M,L$. On the other hand, the neural networks in the proposed method can be learned without knowing these details about the deployment scenario.

\section{Simulations}
\label{sec:simulation}

In Section \ref{sec:nnoptimization}, we proposed a two phase method for solving the class of separable parameters problems: An Offline Phase I followed by an Online Phase II based on ADMM.
The purpose of this section is to illustrate both these phases, using the  motivating problems described in
Section \ref{sec:modelexamples}. We first present simulation
results to illustrate the optimal solution obtained by  Phase II (implementing steps \eqref{eqn:xnopt}, \eqref{eqn:lambdaupdate}, \eqref{eqn:nuupdate}) in these problems, assuming that the precise optimal values $\bfx^\dag,\bff^\dag$ are available. (So, in this part, we do not use neural network representation.) This presentation is for to two reasons. Firstly, since these motivating problems allow more complex variations of classical problems (such as beamforming and waterfilling),  these results will show the value of these generalized problems. Secondly, these results will also show that Phase II converges well, despite the multiple copies $\bfx_{n,m,l}$ made of the variables, and even though its steps only use  additions and multiplications, and don't rely on extensive numerical computation libraries.
Figure \ref{fig:convergence} shows the typical convergence of the ADMM in Phase II; specifically for the Power control problem of Section \ref{sec:modelexamples}. All 10 users had equal SINR threshold $\gamma_n=10$ dB in this example, with $N_0=0.1$. The figure shows that the terminal $SINR$ of each user does meet that threshold, and further that the $(M+1)L$ copies $\bfx_{0,m,l}$ for $n=0$ do eventually become equal. Figure \ref{fig:antennagain} shows the antenna array pattern obtained due to the Robust beam-forming of Section \ref{sec:modelexamples} with a 15-antenna linear array, 4 interferers with $2^o$ angle uncertainties and equal powers $P_l$. The worst-case $SINR$ obtained by the robust beam, given the angle uncertainty about the interferers (shown as red bars), is 32 dB, which is 6.5 dB higher than the standard $SINR$ maximization problem without accounting for angle uncertainty (denoted as `standard beam'), that can be calculated by solving a generalized eigenvalue problem. A closer look at the figure  reveals that the robust beamformer does not put a null at the nominal location of the interferers (to account for the angle uncertainty), allowing it flexibility to increase the main lobe.

\begin{figure}[t]
\includegraphics[width=0.95\columnwidth]{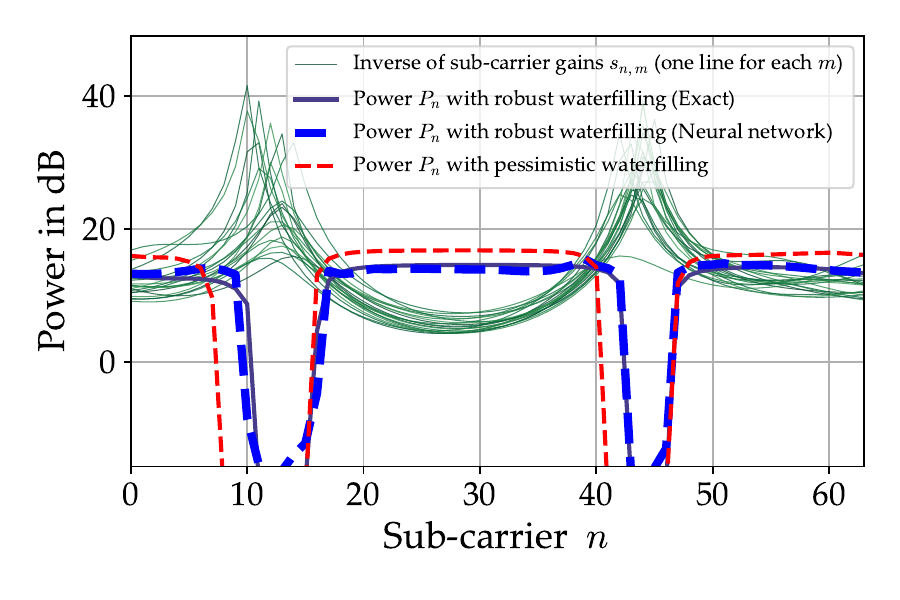}
\vspace{-1em}
\caption{Optimal water-filling under uncertainty.}
\label{fig:waterfillingrobust}
\end{figure}

Turning now to the Offline  Phase I of the proposed method, our results  are currently
preliminary. Note that this offline phase assumes that sufficient computing power is
available, so that the learned neural networks  can
represent the $\bfx^\dag,\bff^\dag$ optimal values accurately. As an example, consider the 
Generalized waterfilling problem \eqref{fig:waterfillingrobust} of Section \ref{sec:modelexamples} with 64 sub-carriers and 4 pilots.  A quadratic objective function $f_{n,0}(x)=x+cx^2$ with $c=2.5\times 10^{-4}$ was chosen here, to penalize any one sub-carrier receiving too much power allocation, and log-capacity expression was used for $r(\cdot)$ with $\gamma_c=0.5,N_0=0.1$. The two $\bfx^\dag$ predictors used 20 layer deep neural networks of width 40   with RELU activation
and an input normalization layer, for a total of 33000 parameters, while the two $\bff^\dag$ predictors had 20 layers of width 20 for a total of 8500 parameters. These were trained using an off-the-shelf GPU (GeForce RTX 3060) in a work computer
using $10^6$ data points.
Figures \ref{fig:xnml} and \ref{fig:fnml}
show a scatter plot of the predictions (represented by hats) made by the {\em x1-Predictor} and {\em f1-Predictor} neural networks 
 versus the true values. The {\em f1-Predictor} is more accurate here (Mean Squared Error of $0.04$ versus $0.01$), even though it uses fewer parameters, and despite each $\bff^\dag$ technically being a function of the corresponding $\bfx^\dag$. 

For this same problem, Figure \ref{fig:waterfillingrobust} shows the optimization result of the proposed method using  both phases;
the Offline phase  learns the neural network representations first, and then the Online phase 
runs ADMM to optimize \eqref{eqn:waterfillingrobust}.
The figure shows that the inverse of channel gains of different sub-carriers are uncertain, due to noise in the measured  sub-carrier gains of the pilot tones. The figure also shows that the optimal solution obtained by the proposed method (called `robust waterfilling (Neural network)' in the plot) selects powers for different sub-carriers that are higher for sub-carriers with higher channel gains, as expected. For comparison, a water-filling type solution designed for the worst case is also shown (marked `pessimistic waterfilling'), which assumes the worst case sub-carrier gain among allowed values. (It is not exactly water-filling because the objective function here is quadratic.) This water-filling solution is worse (incurs higher cost of 2030 versus 1319) compared to the  robust method. This is also clear by noticing that it uses high powers for many sub-carriers compared to the robust method, and thus, is penalized more by the quadratic cost. Finally, it can be seen that running Phase II with exact values of $\bfx^\dag,\bff^\dag$ available gives slightly improved result (see plot marked `Exact') with a cost of 1286. The gap can be closed by using larger neural networks, or perhaps by improving training of these networks.

\section{Conclusions}
\label{sec:conclusions} In this paper, we  proposed a two phase method based on a reformulation of ADMM, to efficiently solve certain parameter-separable parametric constrained optimization problems. The proposed two phase approach combines neural network based function predictors that are trained offline with relatively lightweight online computing, to obtain solutions of instances of the parametric optimization, as parameters vary, on resource limited  hardware. Assuming  accurate function predictors are available, numerical illustrations show the efficacy of the proposed approach with relatively lightweight runtime computing requirements for a number of motivating  problems that are of independent interest. The overall accuracy of the two-phase procedure is naturally limited by the accuracy of the function approximators used, which in turn depends on a number of factors in their offline training phase, e.g., size of the training data and the architecture of the neural network being used. Future work will investigate efficient design of such neural network predictors.

\section{Acknowledgments}
This material is based upon work supported by the US National Science Foundation under Grant 1422193.

\end{document}